\documentclass[letter,11pt]{article}
\pdfoutput=1 % if your are submitting a pdflatex (i.e. if you have
\usepackage{jcappub} % for details on the use of the package, please
\usepackage[normalem]{ulem}

\usepackage[utf8]{inputenc}
\usepackage[T1]{fontenc} % if needed
\usepackage{dsfont}
\usepackage{amsmath,amssymb,calc}
\usepackage{color}
\usepackage{amssymb}
\usepackage{graphicx, epsfig, bm}
\usepackage{hyperref}
\usepackage{soul}
\usepackage{multicol}
\usepackage{changepage}
\usepackage{subfig}
\usepackage{mathtools}
\usepackage{comment}
\usepackage{color}
\usepackage{ifthen}

\def\bdm{\begin{displaymath}}
\def\edm{\end{displaymath}}

\def\barray{\begin{array}}
\def\earray{\end{array}}
\def\be{\begin{equation}}
\def\ee{\end{equation}}
\def\ben{\begin{equation} \nonumber}
\def\een{\end{equation}}
\def\ban{\begin{eqnarray*}}
\def\ean{\end{eqnarray*}}
\def\ba{\begin{eqnarray}}
\def\ea{\end{eqnarray}}
\def\eal{\end{align}}
\def\bal{\begin{align}}

\def\({\left(}
\def\){\right)}
\def\[{\left[}
\def\]{\right]}

\def\by{{\bf{y}}}
\def\ba{{a\left(\by\right)}}

\def\bk{{\bf k}}
\def\bq{{\bf q}}
\def\bx{{\bf x}}
\def\bp{{\bf p}}

\definecolor{gold}{rgb}{1.0, 0.84, 0.0}
\definecolor{maroon}{rgb}{.25,0,0}
\definecolor{darkorange}{rgb}{1.0, 0.55, 0.0}
\definecolor{corn}{rgb}{0.98, 0.93, 0.36}
\definecolor{bronze}{rgb}{0.8, 0.5, 0.2}
\definecolor{darkgreen}{cmyk}{0.85,0.2,1.00,0.2}

\begin{document}

\title{Correlated scalar perturbations and gravitational waves from axion inflation}

\author{Sofia P. Corb\`a}
\affiliation{Amherst Center for Fundamental Interactions, Department of Physics, University of Massachusetts, Amherst, MA 01003, U.S.A.}
\author{and Lorenzo Sorbo}

% The "\note" macro will give a warning: "Ignoring empty anchor..."
% you can safely ignore it.

% e-mail addresses: one for each author, in the same order as the authors

\emailAdd{spcorba@umass.edu}
\emailAdd{sorbo@umass.edu}

\abstract{The scalar and tensor fluctuations generated during inflation can be correlated, if arising from the same underlying mechanism. In this paper we investigate such correlation in the model of axion inflation, where the rolling inflaton produces quanta of a $U(1)$ gauge field which, in turn, source scalar and tensor fluctuations. We compute the primordial correlator of the curvature perturbation, $\zeta$, with the gravitational energy density, $\Omega_{GW}$, at frequencies probed by gravitational wave detectors. This two-point function receives two contributions: one arising from the correlation of gravitational waves with the scalar perturbations generated by the standard mechanism of amplification of vacuum fluctuations, and the other coming from the correlation of gravitational waves with  the scalar perturbations sourced by the gauge field. Our analysis shows that the former effect is generally dominant. For typical values of the parameters, the correlator, normalized by the amplitude of $\zeta$ and by the fractional energy in gravitational waves at interferometer frequencies, turns out to be of the order of $ 10^{-4}\div 10^{-2}$.}

\maketitle


%%%%%%%%%%%%%%%%%%%%%%%
\section{Introduction}%
%%%%%%%%%%%%%%%%%%%%%%%

The theory of inflation constitutes the dominant paradigm of primordial cosmology. Besides solving the most important problems of the standard Hot Big Bang model, it is able to provide an explanation, in excellent agreement with observations, for the origin of the temperature anisotropies present in the Cosmic Microwave Background (CMB) radiation and of the density fluctuations that characterize the large scale structure of the Universe. Among the many different inflationary scenarios, axion inflation is one of those giving a satisfying solution to the problem of UV sensitivity of the inflaton potential. In this model, proposed for the first time in 1990 as natural inflation~\cite{Freese:1990rb}, the inflaton is a pseudo-Nambu--Goldstone Boson that enjoys a (softly broken) shift symmetry, i.e., a symmetry under the transformation $\phi\rightarrow \phi + \text{const}$, which protects its potential against large radiative corrections.

The axionic inflaton is naturally coupled to gauge fields through the operator $\phi F_{\mu\nu}\tilde{F}^{\mu\nu}/f$, where $f$ is the axion decay constant. In the presence of such coupling, the rolling zero mode of the inflaton acts as a source for the modes of the gauge field. As a result, quanta of the gauge field are amplified into classical modes, which in turn source, through a process of inverse decay, both scalar and tensor fluctuations. Since, due to the pseudoscalar nature of the inflaton, only one of the two helicities of the gauge field experiences a tachyonic instability, the spectra of the tensor modes of different helicities have different amplitudes. This scenario has multiple phenomenological predictions, including nongaussianities~\cite{Barnaby:2010vf}, deviations from scale invariance~\cite{Namba:2015gja}, formation of a population of primordial black holes~\cite{Linde:2012bt}, generation of primordial chiral gravitational waves at CMB~\cite{Sorbo:2011rz} or interferometer~\cite{Cook:2011hg} frequencies, baryogenesis~\cite{Anber:2015yca}, as well as the possible generation of cosmologically relevant magnetic fields~\cite{Garretson:1992vt,Anber:2006xt} - see~\cite{Pajer:2013fsa} for a review.

By comparing these phenomenological predictions with observations we can constrain the relevant parameters characterizing the models of axion inflation. More specifically, there are two significant observational lengthscales. At large scales, probed by CMB measurements, the primary constraint arises from the non-observation of primordial nongaussianities for the scalar fluctuations. In axion inflation the sourced scalar fluctuations are highly nongaussian. Consequently, the model can be viable only if the sourced component of scalar modes is subdominant compared to  that generated by the standard amplification of vacuum fluctuations. This is equivalent to stating that the amplitude of the gauge field, which sources the scalar and tensor fluctuations, must be relatively small. Therefore, the sourced component of tensor fluctuations is also small at this stage. 

At smaller scales, corresponding to modes that left the horizon closer to the end of inflation, the situation becomes more interesting. For simple inflationary potentials, the inflaton's velocity increases as inflation progresses and therefore the population of gauge quanta, whose amplitude depends exponentially on the inflaton's velocity, becomes more sizable towards the end of inflation. As a consequence, sourced gravitational waves of shorter wavelengths, which are remarkably those probed by gravitational  wave experiments, can have a much larger amplitude and might even be directly detectable~\cite{Cook:2011hg} by a variety of observatories. Also in this regime we need the scalar fluctuations to remain  bounded to avoid an overproduction of primordial black holes~\cite{Pajer:2013fsa,Garcia-Bellido:2016dkw}.

A natural follow-up to the recent observational evidence~\cite{NANOGrav:2023gor,EPTA:2023fyk,Reardon:2023gzh} of a stochastic gravitational wave background (SGWB) is the search for anisotropies, in analogy to the scalar anisotropies observed in the CMB (see, e.g.,~\cite{KAGRA:2021mth} for a recent analysis of LIGO/Virgo/KAGRA and~\cite{LISACosmologyWorkingGroup:2022kbp} for LISA’s reach in this respect). Study of these anisotropies can allow us to distinguish between the astrophysical and cosmological origin of the SGWB.  Furthermore, cosmological tensor anisotropies may be correlated with the scalar anisotropies of the CMB if they arise from the same underlying mechanisms~\cite{Geller:2018mwu}. Exploring such correlations can give important information about the cosmological background of gravitational waves, thus providing insights about the physics of the Early Universe. Reference~\cite{Schulze:2023ich} performed a study of the statistics of these anisotropies while~\cite{Adshead:2020bji} studied the consequences of a non-trivial primordial scalar-tensor-tensor nongaussianity on the energy density of gravitational waves.

In this work we compute the correlation between the curvature perturbation $\zeta(\bx)$ and the energy density $\Omega_{GW}(\bx)= \dot{h}_{ij}(\bx)\,\dot{h}_{ij}(\bx)/(12\,H_0^2) $ of the tensor modes  within the framework of axion inflation. The computation is conducted at frequencies tested by gravitational detectors, and the correlator is normalized by both the square root of the scalar power spectrum and the average value of $\Omega_{GW}(\bx)$. The two point function receives two contributions, reflecting the fact that scalar fluctuations are generated both from the vacuum, through the standard amplification process, and by  modes of the gauge field, through the inverse decay process. More specifically, we will study the two following situations:

\begin{itemize}	

\item the rolling inflaton has fluctuations that are generated by the standard mechanism of amplification of vacuum fluctuations in an expanding Universe. The rolling inflaton then sources quanta of the gauge field, which in turn source gravitational waves.  The fluctuations in the inflaton are thus imprinted in the fluctuations in the gravitational waves. We study this correlator in Section~\ref{subsec:amplified};
 
\item the rolling inflaton sources quanta of the gauge field, which in  turn source {\em both} scalar fluctuations and gravitational waves. Since these modes are produced by the same population of gauge modes, they are correlated. We study this correlator in Section~\ref{subsec:sourced}.
	
\end{itemize}

As we will see, due to the smallness of the amplitude of the sourced gravitational waves at CMB scales, the former effect is generally dominant over the latter, and leads to a normalized correlator of the order of $10^{-4}\div 10^{-2}$.

The correlator studied in this work is the one between scalar perturbations at CMB scales, corresponding to modes that left the horizon early during inflation and gravitational waves at interferometer scales, which correspond to modes that left the horizon later during inflation. Even though these gravitational waves have relatively short (i.e., non cosmological) wavelengths, their {\em anisotropies} are at large, cosmological scales.

During the last stages of axion inflation the large amplitude acquired by the gauge modes  implies that they can have strong backreaction effects on the inflating background. The nonperturbative inflaton-gauge field dynamics, studied in numerous papers including~\cite{Cheng:2015oqa,Notari:2016npn,Sobol:2019xls,DallAgata:2019yrr,Domcke:2020zez,Caravano:2022epk,Peloso:2022ovc,Figueroa:2023oxc,Garcia-Bellido:2023ser,vonEckardstein:2023gwk,Caravano:2024xsb}, is rich, complicated, and not yet fully understood. The production of gravitational waves, although generated during the phase of strong backreaction, is treated at the perturbative level. Reference~\cite{Garcia-Bellido:2023ser} derived spectra of gravitational waves produced during this stage keeping into account the nonperturbative dynamics of the inflaton-gauge field system, even if it ignored inflaton inhomogeneities. Reference~\cite{Iarygina:2023mtj} performed an analogous study for the case of an $SU(2)$ gauge sector. The results of~\cite{Garcia-Bellido:2023ser} suggest that, even though strong backreaction effects complicate significantly the dynamics of the inflaton and of the gauge quanta, if the inflaton evolution $\phi(t)$ is known, then the resulting gravitational wave spectra reflect quite accurately the shape of the function $\dot\phi(t)$. For the scope of our calculation, since we will formulate our results in terms of $\dot\phi(t)$ without referring to the specific dynamics that led to that expression, our results should be  valid even in the strong backreaction regime, at least as long as the inflaton inhomogeneities are ignored.  Moreover, there are reasons to expect that our results will not change even once inflaton gradients are accounted for, since causality will prevent the late strong dynamics from affecting physics at scales that have left the horizon at much earlier times.

This paper is organized as follows. Section~\ref{sec:background} contains a review of the amplification process that quanta of gauge field undergo as the inflaton rolls down its potential, together with the generation of curvature perturbations and of gravitational waves. Then, in Section~\ref{sec:correlator}, we calculate the two contributions to the correlator between scalar fluctuations and the energy density of the  gravitational waves: in Subsection~\ref{subsec:amplified} we study the correlation of gravitational waves with the amplified vacuum scalar fluctuations and in Subsection~\ref{subsec:sourced} the correlation of gravitational waves with sourced scalar fluctuations. In Section~\ref{sec:conclusion} we discuss our results and we conclude. Appendix~\ref{appendix} contains the details of the calculation leading to the results in Section~\ref{subsec:sourced}.

%%%%%%%%%%%%%%%%%%%%%%%%%%%%%%%%%%%%%%%%%%%%%%%%%%%%%%%%%%%%
\section{Review of scalar and tensor perturbations from axion inflation}
\label{sec:background}%
%%%%%%%%%%%%%%%%%%%%%%%%%%%%%%%%%%%%%%%%%%%%%%%%%%%%%%%%%%%%

Our system consists of a pseudoscalar inflaton $\phi$ and a $U(1)$ gauge field $A_\mu$ in interaction with each other and with gravity through the action 
\begin{align}
{\cal S}=\int d^4x\sqrt{-g}\left[\frac{M_P^2}{2}R-\frac12\partial_\mu\phi\,\partial^\mu\phi-V(\phi)-\frac14F_{\mu\nu}\,F^{\mu\nu}-\frac{\phi}{8\,f}\frac{\epsilon^{\mu\nu\rho\lambda}}{\sqrt{-g}}F_{\mu\nu}\,F_{\rho\lambda}\right]\,, 
\end{align}
where $g={\rm{det}}(g_{\mu\nu})$, $F_{\mu\nu}=\partial_\mu A_\nu-\partial_\nu A_\mu$, $f$ is a constant with dimensions of mass, $R$ is the Ricci scalar, and $\epsilon^{\mu\nu\rho\lambda}$ is the totally antisymmetric object defined by $\epsilon^{0123}=+1$. We will not make any assumption about the shape of the potential $V(\phi)$, other than it is flat enough to be able to support inflation.

Concerning the metric, we will assume that it is of the form of de Sitter space in flat slicing plus tensor perturbations (repeated latin indices are understood to be summed upon)
\begin{align}\label{eq:metric}
ds^2&=a^2(\tau)\left[-d\tau^2+\left(\delta_{ij}+h_{ij}(\bx,\,\tau)\right)\,dx^i\,dx^j\right]\,,\nonumber\\
&\quad a(\tau)=-\frac{1}{H\,\tau}\,,\qquad  h_{ii}=\partial_ih_{ij}=0   \,.
\end{align}
We perturb the inflaton as
\begin{align}
\phi(\bx,\,\tau)\equiv\phi_0(\tau)+\delta\phi(\bx,\,\tau)\,,     
\end{align}
so that the curvature perturbation is given by $\zeta\equiv - \frac{H}{\dot{\phi}_0}\delta\phi$. We will denote the derivative with respect to conformal time $\tau$ by a prime and that with respect to the cosmic time $t$, defined through $dt=a(\tau)\,d\tau$, by an overdot. We set the scale factor to be equal to unity at the end of inflation, i.e., inflation will end at $\tau_{\rm e}=-1/H$.

We treat the homogeneous inflaton $\phi_0(\tau)$ and the scale factor $a(\tau)$ as background quantities, and we  work with the following canonically normalized perturbations 
\begin{align}\label{eq:can_norm_quantities}
&A_\mu(\bx,\,\tau) \, \qquad\text{with}\qquad \,A_0(\bx,\,\tau)=0\,, \quad  \partial_i A_i(\bx,\,\tau)=0\,,\nonumber\\
&\Phi(\bx,\,\tau)\equiv a(\tau)\,\delta\phi(\bx,\,\tau)\,,\nonumber\\
&H_{ij}(\bx,\,\tau)\equiv\frac{M_P}{2}\,a(\tau)\,h_{ij}(\bx,\,\tau)\,.
\end{align}
Neglecting the mass of the inflaton, our perturbed Lagrangian takes the form
\begin{align}\label{eq:quad_lag}
{\cal L}=&\left(\frac{1}{2}\Phi'{}^2-\frac{1}{2}\partial_k\Phi\,\partial_k\Phi+\frac{a''}{2\,a}\Phi^2\right)+\left(\frac12{H}'_{ij}\,{H}'_{ij}-\frac12\partial_k{H}_{ij}\,\partial_k{H}_{ij}+\frac{a''}{2\,a}{H}_{ij}\,{H}_{ij}\right)\nonumber\\
&+\left(\frac12A_i'\,A_i'-\frac12\partial_kA_i\,\partial_kA_i-\frac{\phi_0}{f}\epsilon^{ijk}\,A_i'\,\partial_jA_k\right)\nonumber\\
&-\frac{H_{ij}}{a\,M_P}\,\left[A_i'\,A_j'-\left(\partial_iA_k-\partial_kA_i\right)\left(\partial_jA_k-\partial_kA_j\right)\right]-\frac{\Phi}{f\,a}\epsilon^{ijk}\,A_i'\,\partial_jA_k\,,
\end{align}
where the first line describes the free scalar and free tensor perturbations, the second line describes the free gauge field modes, and the last line contains the interactions that lead to processes of the form $A_iA_j\to H_{ij}$ and $A_iA_j\to \Phi$.

By varying the Lagrangian \eqref{eq:quad_lag} with respect to $\Phi$, $H_{ij}$ and $A_i$, we obtain the equations of motion
\begin{align} \label{eq:eom_Phi}
&\Phi'' - \frac{a''}{a}\Phi- \nabla^2\Phi+\frac{1}{f\,a}  \epsilon^{ijk}\,A_i'\,\partial_jA_k=0\,,\\
\label{eq:eom_H}
&H''_{ij}- \frac{a''}{a} H_{ij} -\nabla^2 H_{ij} +\frac{1}{a\,M_P}\,\left[A_i'\,A_j'-\left(\partial_iA_k-\partial_kA_i\right)\left(\partial_jA_k-\partial_kA_j\right)\right]\,=0\,,\\
\label{eq:eom_A}
&A_i'' -\nabla^2 A_i -\frac{\phi'_0}{f}\epsilon^{ijk}\,\partial_jA_k=0\,.
\end{align}

The solution of eq.~\eqref{eq:eom_Phi} splits into two parts: the solution of the homogeneous equation, denoted as $\Phi_{\rm V}$, and the particular solution, denoted as $\Phi_{\rm S}$. The solution of the homogeneous equation represents the usual vacuum fluctuations generated during inflation due to the accelerated expansion of the background, while the particular solution is induced by the inverse decay of the gauge fields. The homogeneous solution can be quantized through the standard quantization of the free Lagrangian, using the first line of eq.~(\ref{eq:quad_lag}), as
\begin{align}\label{eq:quant_Phi}
\Phi_{\rm V}(\bx,\,\tau)&=\int\frac{d\bk}{(2\pi)^{3/2}}e^{i\bk\bx}\,\left[\Phi_{\rm V}(k,\,\tau)\,\hat{a}(\bk)+\Phi_{\rm V}^*(k,\,\tau)\,\hat{a}^\dagger(-\bk)\right]\,,\nonumber\\
\Phi_{\rm V}(k,\,\tau)&\equiv\frac{1}{\sqrt{2k}}\left(1-\frac{i}{k\tau}\right)\,e^{-ik\tau}\,,
\end{align}
where the creation/annihilation operators $\hat{a}^\dagger(\bk)/\hat{a}(\bk)$ satisfy the usual commutation relations $\left[\hat{a}(\bk),\,\hat{a}^\dagger(\bq)\right]=\delta(\bk-\bq)$, $\left[\hat{a}(\bk),\,\hat{a}(\bq)\right]=\left[\hat{a}^\dagger(\bk),\,\hat{a}^\dagger(\bq)\right]=0$.

The power spectrum of the curvature perturbation, ${\cal P}_\zeta$, defined through the two point function
\begin{align}
\langle \zeta(\bk)\,\zeta(\bq)\rangle\equiv \frac{2\pi^2}{k^3}\,{\cal P}_\zeta(\bk)\,\delta(\bk+\bq)\,,
\end{align}
results in the sum of the power spectra corresponding to the homogeneous and the particular solutions, denoted as ${\cal P}_{\zeta,\,{\rm V}}$ and ${\cal P}_{\zeta,\,{\rm S}}$, respectively.

Specifically, the homogeneous solution, corresponding to the scalar perturbations associated to the mode functions~(\ref{eq:quant_Phi}), yields, at the end of inflation and for large scales, 
\begin{align}
{\cal P}_{\zeta,\,{\rm V}}=\frac{k^3}{2\pi^2}\frac{H^2}{\dot{\phi}_0^2}\left|\Phi_{\rm V}(k,\,\tau_{\rm e})\right|^2\xrightarrow[k\ll H]{} \frac{H^4}{4\pi^2\,\dot{\phi}_0^2}\,.
\end{align}
An analogous discussion holds also for the tensor perturbations $H_{ij}(\bx,\,\tau)$, whose vacuum component gives rise to ${\cal P}_{h,\,{\rm V}} =\frac{2\,H^2}{\pi^2\,M_P^2}$.

In order to find the sourced components of the scalar and tensor power spectra we need to take into account the generation of the electromagnetic field by the rolling pseudoscalar. In order to do that, we start with the quantization of the vector field $A_i(\bx,\,\tau)$:
\begin{align}\label{eq:quantA}
A_{i}(\bx,\,\tau)&=\int\frac{d\bk}{(2\pi)^{3/2}}\sum_{\lambda=\pm} e^{\,\lambda}_{i}(\widehat\bk)\,e^{i\bk\bx}\left[A_\lambda(k,\,\tau)\,\hat{a}_\lambda(\bk)+A_\lambda^*(k,\,\tau)\,\hat{a}^\dagger_\lambda(-\bk)\right]\,,
\end{align}
where the helicity projectors $e^{\pm}_{i}(\widehat\bk)$ satisfy the relations
\begin{align}
\begin{array}{ll}
k_i\,e_i^{\,\lambda}(\widehat{\bk})=0\,,&e_i^{\,\lambda}(\widehat\bk)^*=e_i^{-\lambda}(\widehat\bk)=e_i^{\,\lambda}(-\widehat\bk)\,,\\
i\epsilon_{ijk}k_je_k^{\,\lambda}(\widehat\bk)=\lambda\, k\, e_i^{\,\lambda}(\widehat\bk)\,,\qquad\qquad & e^{\,\lambda}_i(\widehat\bk)e_i^{\,\lambda'}(\widehat\bk)=\delta_{\lambda,\,-\lambda'}\,.
\end{array}
\end{align}
Inserting the decomposition~\eqref{eq:quantA} into eq.~\eqref{eq:eom_A} we obtain the equation of motion for the mode functions $A_\lambda(k,\,\tau)$,
\begin{align}\label{eq:eomAlambda}
A_\lambda''(k,\,\tau)+\left(k^2-\lambda\frac{\phi'_0}{f}k\right)\,A_\lambda(k,\,\tau)=0\,, 
\end{align}
which can be solved explicitly in terms of special functions if $\dot{\phi}_0=$ constant. However, we do not need the exact solution. Defining
\begin{align}
\xi\equiv \frac{\dot{\phi}_0}{2\,f\,H}\,,
\end{align}
we can rewrite eq.~(\ref{eq:eomAlambda}) as
\begin{align}\label{eq:eomAlambda1}
\frac{d^2\,A_\lambda}{d(k\,\tau)^2}+\left(1+2\,\lambda\,\frac{\xi}{k\,\tau}\right)\,A_\lambda=0\,, 
\end{align}
so that, assuming $\xi>0$, the helicity $\lambda=-1$ in eq.~(\ref{eq:eomAlambda1}) has always real frequencies that are adiabatically evolving (remember that $\tau<0$). As a consequence, the mode $A_-$ stays in its vacuum and we will neglect it from now on. On the other hand, the positive helicity mode $A_+$ has imaginary frequencies for a range of values of $k\tau$ and is therefore exponentially amplified.

In the WKB approximation, the leading term in the solution of the tachyonic modes of $A_+$ reads~\cite{Anber:2006xt}
\begin{align}\label{eq:a+appr}
A_+(k,\,\tau)\simeq \frac{1}{\sqrt{2\,k}}\,\left(-\frac{k\,\tau}{2\,\xi}\right)^{1/4}\,e^{-2\,\sqrt{-2\xi k\tau}+\pi\,\xi}\,,
\end{align}
which is strictly speaking valid only in the range~\cite{Barnaby:2010vf} $\frac{1}{8\,\xi}\lesssim |k\,\tau|\lesssim  2\,\xi$ (we will assume $\xi\gtrsim O(1)$ throughout this paper). However, since the momenta in this range dominate the contributions to the observables we will be interested in, we will apply the expression~(\ref{eq:a+appr}) to the entire range $0<|k\,\tau|<\infty$. Eq.~(\ref{eq:a+appr}) shows that the $\lambda=+$ helicity of the gauge field is amplified by a factor $e^{\pi\xi}$, which can be very large even for moderate values of $\xi$.

\smallskip

We are now in position to compute  the leading order contribution of the amplified gauge field to the curvature perturbation $\zeta$. Taking the Fourier of eq.~\eqref{eq:eom_Phi}, we obtain the equation
\begin{align}
&\Phi''(\bq,\,\tau)+q^2 \Phi(\bq,\,\tau)-\frac{2}{\tau^2}\Phi(\bq,\,\tau)-i\,\frac{H\tau}{f}\,\epsilon^{ijk}\int\frac{d\bp}{(2\pi)^{3/2}} \,A_i'(\bp,\,\tau)\,(\bq-\bp)_jA_k(\bq-\bp,\,\tau)=0\,.
\end{align}
The particular solution of this equation, $ \Phi_{\mathrm S}$, which corresponds to the sourced component of scalar fluctuations, can be found using the retarded propagator
\begin{align}\label{eq:phi_sourced}
&\Phi_{\mathrm S}(\bq,\,\tau)\equiv i\,\int d\tau'\,G_q(\tau,\,\tau')\frac{H\tau'}{f}\,\epsilon^{ijk}\int\frac{d\bp}{(2\pi)^{3/2}} \,A_i'(\bp,\,\tau')\,(\bq-\bp)_jA_k(\bq-\bp,\,\tau')\,.
\end{align}
Given that we are assuming an exact de Sitter background, the retarded propagator can be written explicitly as
\begin{align}\label{eq:prop}
G_k(\tau,\tau')= \frac{(1+k^2\,\tau\,\tau')\,\sin(k\,(\tau-\tau'))+k\,(\tau'-\tau)\,\cos(k\,(\tau-\tau'))}{k^3\,\tau\,\tau'}\,\Theta(\tau-\tau')\,,
\end{align}
where $\Theta$ denotes the Heaviside step function.

The sourced component of the scalar fluctuations induces an additional contribution to the power spectrum of the curvature perturbation, that for $\xi\gtrsim 3$, is well approximated by the formula~\cite{Barnaby:2010vf}
\begin{align}
{\cal P}_{\zeta,\,{\rm S}}=\frac{k^3}{2\pi^2}\frac{H^2}{\dot{{\phi}}_0^2}\,\left|\Phi_{\rm S}(k,\,\tau_{\rm e})\right|^2\xrightarrow[k\ll H]{} 4.8\times 10^{-8}\,\frac{H^8}{\dot\phi_0^4}\,\frac{e^{4\pi\xi}}{\xi^6}\,.
\end{align}

A commonly used measure of nongaussianity is the parameter $f_{\rm NL}$, which measures the amplitude of the bispectrum of the curvature perturbation and is defined via
\begin{align}
\langle \zeta(\bk_1)\,\zeta(\bk_2)\,\zeta(\bk_3)\rangle=\frac{3}{10}\left(2\pi\right)^{5/2}\,f_{\rm NL}(k_1,\,k_2,\,k_3)\,{\cal P}_\zeta^2\,\delta(\bk_1+\bk_2+\bk_3)\,\frac{k_1^3+k_2^3+k_3^3}{k_1^3\,k_2^3\,k_3^3}\,.
\end{align}

For single field, slow-roll inflation, the bispectrum has a small amplitude, and $f_{\rm NL}$ is of the order of the slow-roll parameters~\cite{Maldacena:2002vr}. On the other hand, the sourced component of the curvature perturbation, since it results from a $2\to 1$ process, obeys an intrinsically nongaussian statistics. Since such nongaussianities originate from some sub-horizon dynamics, the bispectrum is peaked on equilateral configurations, i.e., for $k_1=k_2=k_3$, with~\cite{Barnaby:2010vf}
\begin{align}\label{eq:fNL}
f_{\rm NL}^{\rm equil}\simeq 7.1 \times 10^{5}\,\frac{H^{12}}{\dot\phi^6}\,\frac{e^{6\pi\xi}}{\xi^9}\,,
\end{align}
for $\xi\gtrsim 3$ and in the regime  ${\cal P}_{\zeta,\,{\rm S}}\ll {\cal P}_{\zeta,\,{\rm V}}$. In the regime of large $\xi$, where ${\cal P}_{\zeta,\,{\rm S}}\gg {\cal P}_{\zeta,\,{\rm V}}$, $f_{\rm NL}^{\rm equil}$ converges to a value of the order of $10^4$, which exceeds by a $O(10^3)$ factor the constraints from Planck. This limits severely the value $\xi_{\rm CMB}$ taken by $\xi$ when Cosmic Microwave Background scales are leaving the horizon, leading to $\xi_{\rm CMB}\lesssim 2.5$~\cite{Barnaby:2011vw,Planck:2015sxf}.

The excited modes of the vector field are also a source of gravitational waves. To leading order, production of gravitational waves via this process is described by the equation
\begin{align}\label{eq:hij_sourced}
&H_{ij}''(\bq,\,\tau)+q^2 H_{ij}(\bq,\,\tau)-\frac{2}{\tau^2}H_{ij}(\bq,\,\tau)\nonumber\\
&\qquad\qquad =\frac{H\,\tau}{M_P}\int\frac{d\bp}{(2\pi)^{3/2}}\left(A_i'(\bp,\,\tau)\,A_j'(\bq-\bp,\,\tau)-F_{ik}(\bp,\,\tau)F_{jk}(\bq-\bp,\,\tau)\right)\,,
\end{align}
where $F_{ij}(\bp,\,\tau)\equiv i\,\bp_iA_j(\bp,\tau)-i\,\bp_jA_i(\bp,\tau)$. As a consequence of the functional dependence of $A_+$ on $k\,\tau$ and on $\xi$, the electric field is stronger than the magnetic field by a factor $\sim \xi\gtrsim 1$. For this reason we will neglect the term $F_{ik}(\bp,\,\tau)F_{jk}(\bq-\bp,\,\tau)$ in eq.~(\ref{eq:hij_sourced}). Using again the Green's function~(\ref{eq:prop}) we eventually obtain     	
\begin{align}\label{eq:h_sourced}
&H_{ij,\,\mathrm S}(\bq,\,\tau)\equiv \int d\tau' \,G_q(\tau,\tau')\, \frac{H \, \tau'}{M_P} \int \frac{d\bp}{(2\pi)^{3/2}} \,A_i'(\bp,\,\tau')\,A_j'(\bq-\bp,\,\tau')\,.
\end{align}
The resulting power spectrum for the tensor modes reads~\cite{Sorbo:2011rz}
\begin{align}\label{eq:Ph}
{\cal P}_h={\cal P}_{h,\,{\rm V}}+{\cal P}_{h,\,{\rm S}}\simeq \frac{2\,H^2}{\pi^2\,M_P^2}+8.7\times 10^{-8}\frac{H^4}{M_P^4}\,\frac{e^{4\pi\xi}}{\xi^6}\,.
\end{align}
It is worth stressing that the sourced component of the gravitational waves is almost fully chiral, as a consequence of the fact that only the $+$ helicity of the gauge field is excited. While this fact can lead to a rich and interesting phenomenology, we will not be concerned with it here.

The constraint on the parameter $\xi$ coming from the limits on nongaussianities implies that ${\cal P}_{h,\,{\rm V}}\gg {\cal P}_{h,\,{\rm S}}$. This constraint, however, holds only for the value $\xi_{\rm CMB}$ taken by $\xi$ when CMB scales left the horizon. The quantity $\xi$ is slowly evolving, typically increasing, during inflation. Since the sourced component of the gravitational wave spectrum has an exponential dependence on $\xi$, it is possible that at later times ${\cal P}_{h,\,{\rm V}}$ is actually overwhelmed by $ {\cal P}_{h,\,{\rm S}}$. We will denote by $\xi_{\rm INT}>\xi_{\rm CMB}$ the value taken by $\xi$ at this later stage, where the subscript ${}_{\rm INT}$ refers to the fact that we are thinking of frequencies probed by gravitational interferometers. In particular, this leads to the possibility that gravitational waves sourced by the vector field have such large amplitude to be directly detectable by current or future gravitational detectors~\cite{Cook:2011hg}.

In the next section we will describe two mechanisms that induce correlation between the curvature perturbation and the gravitational waves produced in axion inflation.

%%%%%%%%%%%%%%%%%%%%%%%%%%%%%%%%%%%%%%%%%%%%%%%%%%%%%%%%%%%%%%
\section{The correlator between scalar fluctuations and gravitational waves}
\label{sec:correlator}%
%%%%%%%%%%%%%%%%%%%%%%%%%%%%%%%%%%%%%%%%%%%%%%%%%%%%%%%%%%%%%%

We define the normalized correlator of scalar fluctuations and gravitational waves as
\begin{align}\label{eq:new_correl}
{\cal C}_{\Omega\zeta}(\bk,\,t_0)&\equiv\frac{1}{\Omega_{GW}^{\rm INT}\,\sqrt{{\cal P}_\zeta^{\rm CMB}}}\frac{k^3}{2\pi^2}\int d\by\,e^{-i\bk\by}\langle \Omega_{GW}(\bx+\by,\,t_0)\,\zeta(\bx,\,t_0)\rangle\,\nonumber\\
&=\frac{1}{\Omega_{GW}^{\rm INT}\,\sqrt{{\cal P}_\zeta^{\rm CMB}}}\frac{k^3}{2\pi^2} \langle \Omega_{GW}(\bk,\,t_0)\,\zeta(-\bk,\,t_0)\rangle'\,,
\end{align}	
where the symbol $\langle\ldots\rangle'$ denotes the correlator stripped of the Dirac delta associated to momentum conservation and $t_0$ indicates the present value of cosmic time. Moreover, $\Omega_{GW}^{\rm INT}$ denotes the fractional energy in gravitational waves at interferometer frequencies, whereas ${\cal P}_\zeta^{\rm CMB}$ denotes the amplitude of scalar perturbations at CMB scales. Given the weak scale dependence of ${\cal P}_\zeta^{\rm CMB}$, from now on we will drop the index ${}^{\rm CMB}$ from ${\cal P}_\zeta$, and will treat this quantity as constant. 
On the other hand, axion inflation can lead to a strong scale dependence of the energy in gravitational waves, which cannot be ignored in our analysis. 

To proceed we observe that $\Omega_{GW}(\bk)=\frac{1}{12\,H_0^2}\int \frac{d\bp}{(2\pi)^{3/2}}\, |\bk-\bp|\,p\,{h}_{ij}(\bk-\bp,\,t_0){h}_{ij}(\bp,\,t_0)$.  The current amplitude $h_{ij}(k,\,t_0)$ is related to the primordial amplitude calculated at the end of inflation $h_{ij}(k,\,t_e)$ through the transfer function ${\cal T}(k)$, which is proportional to $k^{-1}$ for modes that have re-entered the horizon during radiation domination, and to $k^{-2}$ for modes that have re-entered the horizon during matter domination. Putting everything together, we have
\begin{align}\label{eq:new_correl2}
{\cal C}_{\Omega\zeta}(\bk,\,t_0)=\frac{1}{12\,H_0^2\,\Omega_{GW}^{\rm INT}\,\sqrt{{\cal P}_\zeta}}\frac{k^3}{2\pi^2} \int\frac{d\bp}{(2\pi)^{3/2}}&\hat{\cal T}(|\bk-\bp|)\,\hat{\cal T}(p)\nonumber\\
&\times \langle {h}_{ij}(\bk-\bp,\,t_e)\,{h}_{ij}(\bp,\,t_e)\,\zeta(-\bk,\,t_e)\rangle'\,,
\end{align}	
where we have defined $\hat{\cal T}(p)\equiv p\,{\cal T}(p)$ and we have replaced the amplitude of the scalar perturbations with its value at the end of inflation.

The correlator ${\cal C}_{\Omega\zeta}(\bk,\,t_0)$ receives two different contributions: the first is the result of the correlation of gravitational waves with the amplified vacuum scalar fluctuations; the second is due to the correlation of gravitational waves with the sourced scalar fluctuations. Below we will examine the two cases separately. 

%%%
\subsection{Correlation with amplified vacuum scalar fluctuations}
\label{subsec:amplified}%
%%%%

The spectrum ${\cal P}_{h,\,{\rm S}}$  of gravitational waves sourced by the gauge field depends on the values of $\phi$ and $\dot\phi$ evaluated approximately at the time when the tensor modes under consideration left the horizon, and where, in slow-roll approximation,  $\dot\phi$ is a function of $\phi$. As a consequence, long wavelength perturbations in the values of $\phi$  will lead to correlated long wavelength perturbations in the spectrum of gravitational waves.  

To first order in the vacuum-amplified fluctuation $\delta\phi_{\rm V}$ of the inflaton, and in the limit in which the wavelength of $\delta\phi_{\rm V}$ is much larger than that  of $h_{ij,\,{\rm S}}$, we have 
\begin{align}
h_{ij,\,{\rm S}}(\bx,\,\phi(\bx))=h_{ij,\,{\rm S}}(\bx,\,\phi_0)+\frac{\partial h_{ij,\,{\rm S}}(\bx,\,\phi_0)}{\partial\phi_0}\,\delta\phi_{\rm V}(\bx)\,,
\end{align}
where the first term does not contribute to ${\cal C}_{\Omega\zeta}$. Since $h_{ij,\,{\rm S}}(\bx,\,\phi_0)\propto e^{2\pi\xi}$, we can also write
\begin{align}
h_{ij,\,{\rm S}}(\bx,\,\phi(\bx))=h_{ij,\,{\rm S}}(\bx,\,\phi_0)\left(1-2\pi\frac{d\,\xi}{d\phi_0}\,\frac{\dot\phi_0}{H}\,\zeta_{\rm V}(\bx)\right)\,,
\end{align}
where we used $\delta\phi=-\dot\phi_0\,\zeta/H$.
We thus obtain the first contribution to the correlator between $\Omega_{GW}$ and $\zeta_{\rm V}$, that we denote as $({\cal C}_{\Omega\zeta})_{\rm V}$, and which reads
\begin{align}\label{eq:modulated_sp}
({\cal C}_{\Omega\zeta})_{\rm V}=&-\frac{1}{12\,H_0^2\,\Omega_{GW}^{\rm INT}\,\sqrt{{\cal P}_\zeta}}\frac{k^3}{2\pi^2}\, \int\frac{d\bp\,d\bq}{(2\pi)^{3}}\hat{\cal T}(|\bk-\bp|)\,\hat{\cal T}(|\bp-\bq|)\nonumber\\
&\times 4\pi\,\frac{\dot\phi_0}{H}\,\frac{d\,\xi}{d\phi_0}\langle {h}_{ij,\,{\rm S}}(\bk-\bp,\,t_e)\,{h}_{ij,\,{\rm S}}(\bp-\bq,\,t_e)\,\zeta_{\rm V}(\bq,\,t_e)\,\zeta_{\rm V}(-\bk,\,t_e)\rangle'\,.
\end{align}	
Assuming $\dot\phi_0>0$, $V'<0$, we have
\begin{align}
\xi\equiv\frac{\dot\phi_0}{2\,fH}\simeq -\frac{V'}{6\,fH^2}=-\frac{M_P^2}{2\,f}\frac{V'}{V}\,,
\end{align}
so that
\begin{align}
\frac{d\xi}{d\phi_0}=-\frac{M_P^2}{2\,f}\left(\frac{V''}{V}-\frac{V'{}^2}{V^2}\right)=\left(\epsilon-\frac{\eta}{2}\right)\frac{1}{f}\,,
\end{align}
where we have defined as usual the slow-roll parameters as
\begin{align}
\epsilon=\frac{M_P^2}{2}\,\frac{V'{}^2}{V^2}\,,\qquad\qquad \eta=M_P^2\frac{V''}{V}\,.
\end{align}
The correlator therefore becomes
\begin{align}
({\cal C}_{\Omega\zeta})_{\rm V}=-\frac{\sqrt{{\cal P}_{\zeta}}}{12\,H_0^2\,\Omega_{GW}^{\rm INT}}\,\int\frac{d\bp}{p^3}\,\xi\,\left(2\epsilon-\eta\right)\,\hat{\cal T}(p)^2\,{\cal P}_{h,\,{\rm S}}(p)\,.
\end{align}

To proceed we note that, since typically the amplitude of the induced tensor modes increases as inflation progresses, the integral in eq.~(\ref{eq:modulated_sp}) is dominated by the largest frequencies, that are typically close to those probed by the interferometers. For those wavelengths, that re-entered the horizon well into the radiation dominated regime, we have
\begin{align}
\frac{\hat{\cal T}(p)^2\,{\cal P}_{h,\,{\rm S}}(p)}{12\,H_0^2\,\Omega_{GW}^{\rm INT}}=\frac{{\cal P}_{h,\,{\rm S}}(p)}{{\cal P}_{h,\,{\rm S}}(p^{\rm INT})}\,.
\end{align}
Using again the fact that the integral in eq.~(\ref{eq:modulated_sp}) is dominated by values of $p$ of the order of $p^{\rm INT}$, we can estimate
\begin{align}\label{eq:fin_vacuum}
({\cal C}_{\Omega\zeta})_{\rm V}\simeq -4\pi\,\xi \,\Delta{\cal N}_*\,(2\epsilon-\eta)\,\sqrt{{\cal P}_\zeta}\,,
\end{align}
where both $\xi$ and the slow-roll parameters $\epsilon$ and $\eta$ are evaluated at the time when the scales probed by interferometers  have left the horizon. In eq.~(\ref{eq:fin_vacuum}) the parameter $\Delta{\cal N}_*$ accounts for the number of efoldings during which the tensor power spectrum is approximately constant. Numerical simulations indicate that this is the case in the strong backreaction regime, which usually lasts $\Delta{\cal N}_*\simeq 10\div 30$ efoldings.
At this stage the parameter $\xi$ takes values that are typically of the order of $5 \div 10$. The quantity $(2\epsilon-\eta)$ has to be smaller than unity and is typically of the order of $10^{-2}\div 10^{-1}$. So by putting everything together we obtain that $({\cal C}_{\Omega\zeta})_{\rm V}$ is typically of the order of $10^{-4}\div 10^{-2}$.

%%%%
\subsection{Correlation with sourced scalar fluctuations}
\label{subsec:sourced}%
%%%

In order to calculate the correlator between the sourced scalar and tensor fluctuations, that we denote as $\left({\cal C}_{\Omega\zeta}\right)_{\rm S}$, we use eqs.~\eqref{eq:can_norm_quantities},~\eqref{eq:phi_sourced} and~\eqref{eq:h_sourced} to find $ \langle h_{ab,\,\mathrm S}(\bk_1,\tau)\,h_{ab,\,\mathrm S}(\bk_2,\tau)\,\zeta_{\mathrm S}(\bk_3,\tau)\rangle$ in terms of the canonically normalized perturbations as 
\begin{align}\label{hhh}
\langle	&h_{ab,\,\mathrm S}(\bk_1,\tau)\,h_{ab,\,\mathrm S}(\bk_2,\tau)\,\zeta_{\mathrm S}(\bk_3,\tau)\rangle= -\frac{4\,H(\tau)}{M_P^2 \,\dot{\phi}_0(\tau)\, a^3(\tau) }\langle H_{ab,\,\mathrm S}(\bk_1,\tau)\,H_{ab,\,\mathrm S}(\bk_2,\tau)\,\Phi_{\mathrm S}(\bk_3,\tau)\rangle \nonumber\\
&= \frac{4\,H(\tau)}{M_P^4\,\dot{\phi}_0(\tau)\,a^3(\tau)\,f } \int^\tau_{-\infty}\, \frac{d\tau_1}{a(\tau_1)}\, \frac{d\tau_2}{a(\tau_2)}\, \frac{d\tau_3}{a(\tau_3)}\, G_{k_1}(\tau,\tau_1)\,G_{k_2}(\tau,\tau_2)\,G_{k_3}(\tau,\tau_3)\nonumber\\  
&\times\int \frac{d\bq_1\,d\bq_2\,d\bq_3}{(2\pi)^{9/2}}\,e^+_a(\widehat{\bq_1})\,e^+_b(\widehat{\bk_1-\bq_1})\,e^+_a(\widehat{\bq_2})\,e^+_b(\widehat{\bk_2-\bq_2})\,e^+_i(\widehat{\bq_3})\,e^+_i(\widehat{\bk_3-\bq_3}) \, |\bk_3-\bq_3| \nonumber \\
&\times\langle A'_+(q_1,\tau_1)\,A'_+(|\bk_1-\bq_1|,\tau_1)\,A'_+(q_2,\tau_2)\,A'_+(|\bk_2-\bq_2|,\tau_2)\,A'_+(q_3,\tau_3)\,A_+(|\bk_3-\bq_3|,\tau_3)\rangle \,,
\end{align}
where we have assumed that only the positive helicity photons contribute because, from eq.~(\ref{eq:eomAlambda}), $A_+$ is the only helicity that is amplified.

Using Wick's theorem to decompose the last line of eq.~\eqref{hhh} and inserting it back into~\eqref{eq:new_correl2}  we obtain
\begin{align}\label{eq:Comegazeta2}
&\left({\cal C}_{\Omega\zeta}\right)_{\rm S}=\frac{k^3\,H(\tau)}{6\,H_0^2\,\pi^2\,M_P^4\,\dot{\phi}_0(\tau)\,a^3(\tau)\,f\,\Omega_{GW}^{\rm INT}\,\sqrt{{\cal P}_\zeta}}\int \frac{d\bp}{(2\pi)^{3/2}} \int^\tau_{-\infty}\, \frac{d\tau_1}{a(\tau_1)}\,\frac{d\tau_2}{a(\tau_2)}\,\frac{d\tau_3}{a(\tau_3)} \nonumber \\
& \times G_{k_1}(\tau,\tau_1)\,G_{k_2}(\tau,\tau_2)\,G_{k_3}(\tau,\tau_3)
\int \frac{d\bq}{(2\pi)^{9/2}}\, \hat{\cal T}(|\bk-\bp|)\,\hat{\cal T}(p)\,{\cal A}(\bq,\bk_1-\bq,\bk_2+\bq)\nonumber\\
&\times \biggl(|\bk_2+\bq|\, A'_+(q,\tau_1)\,A_+'(|\bk_1-\bq|,\tau_1)	\,A_+'(q,\tau_2)\,A_+'(|\bk_1-\bq|,\tau_3)\,A_+'(|\bk_2+\bq|,\tau_2)\,A_+(|\bk_2+\bq|,\tau_3)\nonumber\\
&+|\bk_1-\bq|\, A'_+(q,\tau_1)\,A_+'(q,\tau_2)\,A_+'(|\bk_1-\bq|,\tau_1)\,A_+(|\bk_1-\bq|,\tau_3)\,A_+'(|\bk_2+\bq|,\tau_2)\,A_+'(|\bk_2+\bq|,\tau_3)\biggr)\,,
\end{align}
where $\bk_1= \bk-\bp$, $\bk_2=\bp$ and $\bk_3= -\bk$, and where we have collected the angular part into the expression~${\cal A}$:
\begin{equation*}
{\cal A}(\bk_1,\bk_2,\bk_3)=\delta_{ac}\,\delta_{bd}\, ((e^+_a(\widehat{\bk_1})\,e^+_c(\widehat{-\bk_1})\,e^+_b(\widehat{\bk_2})\,e^+_i(\widehat{-\bk_2})\,e^+_d(\widehat{\bk_3})\,e^+_i(\widehat{-\bk_3})+(a\leftrightarrow b))+(c\leftrightarrow d))\,.
\end{equation*}
Using the explicit form of the gauge field~\eqref{eq:a+appr}, the expression~\eqref{eq:Comegazeta2} becomes
\begin{align}
&\left({\cal C}_{\Omega\zeta}\right)_{\rm S}=-\frac{k^3\,H^4(\tau)}{ 3\times 2^9\,\pi^8\,H_0^2\,M_P^4\,\dot{\phi}_0(\tau)\,f\,a^3(\tau)\,\Omega_{GW}^{\rm INT}\,\sqrt{{\cal P}_\zeta}}\int d\bp\, \hat{\cal T}(|\bk-\bp|)\,\hat{\cal T}(p)\nonumber\\
&\times\int_{-\infty}^\tau d\tau_1\,d\tau_2\,d\tau_3 \,\xi_1^{1/2}\,\xi_2^{1/2}\,\sqrt{\tau_1\,\tau_2} \,\tau_3\,G_{|\bk-\bp|}(\tau,\tau_1)\,G_{p}(\tau,\tau_2)\,G_{k}(\tau,\tau_3)e^{2\pi(\xi_1+\xi_2+\xi_3)}\nonumber\\
&\times\int d\bq\,{\cal A}(\bq,\bk-\bp-\bq,\bp+\bq) 	q^{1/2}\,|\bk-\bp-\bq|^{1/2}\,|\bp+\bq|^{1/2} (|\bk-\bp-\bq|^{1/2}+|\bp+\bq|^{1/2} )\nonumber\\
&\times e^{-2\sqrt{-2\,\xi_1\,q\,\tau_1}-2\sqrt{-2\,\xi_1\,|\bk-\bp-\bq|\,\tau_1}-2\sqrt{-2\,\xi_2\,q\,\tau_2}-2\sqrt{-2\,\xi_2\,|\bp+\bq|\,\tau_2}-2\sqrt{-2\,\xi_3\,|\bp+\bq|\,\tau_3}-2\sqrt{-2\,\xi_3\,|\bk-\bp-\bq|\,\tau_3}}\,,
\,
\end{align}
where we have also allowed for the parameter $\xi$ to be time-dependent, albeit adiabatically, and we have denoted $\xi_i\equiv \xi(\tau_i)$. In order to perform the calculation we set the time at the end of inflation to be $\tau_{\rm e}=-1/H$.  
Since we are interested in modes that are well outside of the horizon at the end of inflation, we will assume $k/H\to 0$.	The dependence of the integrand on $e^{-2\sqrt{-2\,\xi_1\,\tau_1}\left(\sqrt{q}+\sqrt{|\bk-\bp-\bq|}\right)}$ with $\xi_1\gg 1$ implies that we can set $|\bk-\bp|\,|\tau_1|\ll 1$ in the propagator, and we can approximate $G_{|\bk-\bp|}(\tau,\,\tau_1)\simeq -\tau_1^2/(3\,\tau)$. A similar argument applies to the other two propagators which are approximated as $G_{p}(\tau,\,\tau_2)\simeq -\tau_2^2/(3\,\tau)$ and $G_{k}(\tau,\,\tau_3)\simeq -\tau_3^2/(3\,\tau)$. As a consequence, the dependence of the integrand on $\tau_1$, $\tau_2$ and $\tau_3$ takes the form $\tau_1^{5/2}\,e^{-2\sqrt{-2\,\xi_1\,\tau_1}\left(\sqrt{q}+\sqrt{|\bk-\bp-\bq|}\right)+2\pi\xi_1}$, $\tau_2^{5/2}\,e^{-2\sqrt{-2\,\xi_2\,\tau_2}\left(\sqrt{q}+\sqrt{|\bp+\bq|}\right)+2\pi\xi_2}$ and $\tau_3^{3}\,e^{-2\sqrt{-2\,\xi_3\,\tau_3}\left(\sqrt{|\bp+\bq|}+\sqrt{|\bk-\bp-\bq|}\right)+2\pi\xi_3}$ respectively. To proceed with the calculation, we need to know the explicit form of the model-dependent function $\xi(\tau)$.
Without choosing a particular model, we can still estimate the integral by assuming that $\xi$ has a weak dependence on $\tau$.
In this case we see that the integral is dominated by values of $|\tau_1|$, $|\tau_2|$ and $|\tau_3|$ belonging respectively to a relatively narrow window around $(\sqrt{q}+\sqrt{|\bk-\bp-\bq|})^{-2}$, $(\sqrt{q}+\sqrt{|\bp+\bq|})^{-2}$ and $(\sqrt{|\bp+\bq|}+\sqrt{|\bk-\bp-\bq|})^{-2}$. We can therefore approximate 
\begin{align}\label{eq:xi123}
&\xi_1=\xi(\tau\simeq - (\sqrt{q}+\sqrt{|\bk-\bp-\bq|}\,)^{-2})\,,\nonumber\\
&\xi_2=\xi(\tau\simeq - (\sqrt{q}+\sqrt{|\bp+\bq|}\,)^{-2})\,,\nonumber\\
&\xi_3=\xi(\tau\simeq - (\sqrt{|\bp+\bq|}+\sqrt{|\bk-\bp-\bq|}\,)^{-2})\,,
\end{align}
which are now momentum-dependent. Using the expression
\begin{align}
	&\int_0^\infty dx\,x^{n-1}\,e^{-a\,\sqrt{x}}= \frac{2}{a^{2n}}\Gamma(2n)\,,
\end{align}
we obtain 
\begin{align}\label{eq:int_dpdq}
&\left({\cal C}_{\Omega\zeta}\right)_{\rm S}=\frac{k^3\,H^7\,\Gamma(7)^2\,\Gamma(8)}{2^{39}\times 3^4\,H_0^2\,\pi^{8}\,M_P^4\,\dot{\phi}_0(\tau)\,f\,\Omega_{GW}^{\rm INT}\,\sqrt{{\cal P}_\zeta}}\int d\bp\,d\bq	\frac{e^{2\pi(\xi_1+\xi_2+\xi_3)}}{\xi_1^3\,\xi_2^3\,\xi_3^4} \nonumber\\
&\times\hat{\cal T}(|\bk-\bp|)\,\hat{\cal T}(p)
\frac{{\cal A}(\bq,\bk-\bp-\bq,\bp+\bq)\, 	q^{1/2}\,|\bk-\bp-\bq|^{1/2}\,|\bp+\bq|^{1/2}}{(\sqrt{q}+\sqrt{|\bk-\bp-\bq|}\,)^7\,(\sqrt{q}+\sqrt{|\bp+\bq|}\,)^7\,(\sqrt{|\bp+\bq|}+\sqrt{|\bk-\bp-\bq|}\,)^7}\,.
\end{align}

The computation of the remaining six-dimensional integral is complicated, again, by the fact that the function $\xi(\tau)$ is model dependent. Even if the time dependence is weak (i.e., slow-roll implies that $d\xi/dt\ll H\,\xi$), we cannot neglect it, because $\xi$ appears in exponents. Moreover, $\xi$ is in general increasing during inflation. The time- (and therefore $\bp$- and $\bq$-) dependence in the exponent leads the coefficient $e^{2\pi(\xi_1+\xi_2+\xi_3)}$ to be an increasing function of the integration variables. On the other hand,  the factors $(...)^{-7}\times (...)^{-7}\times (...)^{-7}$ in the denominator of eq.~(\ref{eq:int_dpdq}) give a contribution that is peaked at small values of $|\bp|$ and $|\bq| $, i.e. $|\bp|\approx|\bq|\approx k$. The result of the integral will thus depend on whether it is dominated by $|\bp|\approx|\bq|\approx k$ or by the largest values of $|\bp|$ and $|\bq|$.

To proceed with our estimates, we assume that the function $\xi(\tau)$ is monotonically increasing, which, as we said, is what typically happens. It will take a value $\xi(\tau=-1/k)\equiv \xi_k$ when scales with comoving wavenumber $k$, leave the horizon, $N_k$  efoldings before the end of inflation. In particular, we have in mind the case where $k\simeq k_{\rm CMB}$, with $N_{\rm CMB}\simeq 60$ (as noted above, observations constrain $\xi_{\rm CMB}\lesssim 2.5$~\cite{Planck:2015sxf}). At a later time, denoted by $\tau_{\rm BR}$, i.e. $N_{\rm BR}=\log(-H\tau_{\rm BR})$ efoldings before the end of inflation, the system gets into the strong backreaction regime, and $\xi$ takes the value $\xi=\xi_{\rm BR}$. The behavior of the system in this regime is still object of active research, but it is reasonable to assume that $\xi$ will be approximately constant for $\tau>\tau_{\rm BR}$, so that the integral does not receive significant contributions by the values of $\bp$ and $\bq$ corresponding to scales that left the horizon after $\tau_{\rm BR}$.

As we show in the Appendix, the integral is dominated by $|\bp|\approx|\bq|\approx k$ if $\xi_{\rm BR}-\xi_k\lesssim (N_k-N_{\rm BR})/(2\pi)$, and by $|\bp|\approx|\bq|\approx -1/\tau_{\rm BR}$ otherwise. Let us examine these two cases separately.

\subsubsection{$\xi_{\rm BR}-\xi_k\lesssim \frac{N_k-N_{\rm BR}}{2\pi}$}

In this case the integral is dominated by $|\bp|\approx|\bq|\approx k$, so that we can set $\xi_1\simeq \xi_2\simeq \xi_3\equiv \xi_k$ everywhere. Moreover, since we are assuming that $k$ is at CMB scales, it corresponds to wavenumbers that reentered the horizon during matter domination, so that we can assume $\hat{\cal T}(k)\simeq \bar{k}^2/k$, where $\bar{k}^2\equiv \frac{3}{4\sqrt2}\,{k_{\rm eq}\,H_0\,\sqrt{\Omega_{\rm rad}}}\simeq (.5\,H_0)^2$, with $k_{\rm eq}$ being the scale that reentered the horizon during matter-radiation equality~\cite{Caprini:2018mtu}. We are thus left with
\begin{align}
&\left({\cal C}_{\Omega\zeta}\right)_{\rm S}=\frac{H^7\,\Gamma(7)^2\,\Gamma(8)\,\bar{k}^4}{2^{39}\times 3^4\,\pi^{8}\,H_0^2\,M_P^4\,\dot{\phi}_0\,f\,\Omega_{GW}^{\rm INT}\,\sqrt{{\cal P}_\zeta}}\frac{e^{6\pi\xi_k}}{\xi_k^{10}}\nonumber\\
&\times k^3\,\int \frac{d\bp\,d\bq}{p\,|\bk-\bp|}	
\frac{{\cal A}(\bq,\bk-\bp-\bq,\bp+\bq)\, 	q^{1/2}\,|\bk-\bp-\bq|^{1/2}\,|\bp+\bq|^{1/2}}{(\sqrt{q}+\sqrt{\bk-\bp-\bq})^7\,(\sqrt{q}+\sqrt{\bp+\bq})^7\,(\sqrt{|\bp+\bq|}+\sqrt{|\bk-\bp-\bq})^7}\,,
\end{align}
where the integral on the second line can be computed numerically, using
\begin{align}
&{\cal A}(\bk_1,\,\bk_2,\,\bk_3)=\frac{1}{4} \biggl(2+3\,(\widehat{\bk_2}\cdot\widehat{\bk_3})^2 -5\,\widehat{\bk_2}\cdot\widehat{\bk_3}+ (\widehat{\bk_1}\cdot\widehat{\bk_3})^2 +(\widehat{\bk_1}\cdot\widehat{\bk_2})^2-\widehat{\bk_1}\cdot\widehat{\bk_3}+\widehat{\bk_1}\cdot\widehat{\bk_2}
\nonumber\\&
\qquad-(\widehat{\bk_1}\cdot\widehat{\bk_3})(\widehat{\bk_1}\cdot\widehat{\bk_2})- (\widehat{\bk_2}\cdot\widehat{\bk_3})(\widehat{\bk_1}\cdot\widehat{\bk_2})
+(\widehat{\bk_2}\cdot\widehat{\bk_3})(\widehat{\bk_1}\cdot\widehat{\bk_3})	 \nonumber\\
&\qquad-(\widehat{\bk_1}\cdot\widehat{\bk_2})(\widehat{\bk_1}\cdot\widehat{\bk_3})(\widehat{\bk_2}\cdot\widehat{\bk_3})-i\,(\widehat{\bk_1}\cdot\widehat{\bk_2}- \widehat{\bk_1}\cdot\widehat{\bk_3}-\widehat{\bk_2}\cdot\widehat{\bk_3}+1)\,\epsilon_{dil}\,\widehat{\bk_1}_d\,\widehat{\bk_2}_i \,\widehat{\bk_3}_l\,\biggl)\,.
\end{align}

One thus obtains
\begin{align}
&\left({\cal C}_{\Omega\zeta}\right)_{\rm S}(k)\simeq 6\times 10^{-12} \,\frac{H^7\,\bar{k}^2}{H_0^2\,M_P^4\,\dot{\phi}_0\,f\,\Omega_{GW}^{\rm INT}\,\sqrt{{\cal P}_\zeta}}\frac{e^{6\pi\,\xi_k}}{\xi_k^{10}}\,\frac{\bar{k}^2}{k^2}\,.
\end{align}
After substituting  $\sqrt{{\cal P}_\zeta}\simeq \sqrt{{\cal P}_{\zeta,\,{\mathrm V}}}= H^2/(2\pi\,\dot{\phi_0})$  and $\Omega_{GW}^{\rm INT}\simeq\frac{\Omega^0_{\rm rad}}{24}\,{\cal P}_{h,\mathrm S}(k_{\rm INT})$~\cite{Caprini:2018mtu}, with $\Omega^0_{\rm rad} \simeq 8.2\times 10^{-5}$ and ${\cal P}_{h,\mathrm S}(k_{\rm INT})$ from~\eqref{eq:Ph}, we obtain the simple form
\begin{equation}\label{eq:C_final2}
\left({\cal C}_{\Omega\zeta}\right)_{\rm S}\simeq 8\,\frac{H_0^2}{k^2}\frac{H}{f}\,e^{6\pi\,\xi_k-4\pi\,\xi_{\rm INT}}\,\frac{\xi_{\rm INT}^6}{\xi_k^{10}}\,.
\end{equation}
Finally, if $k$ is at CMB scales, we use eq.~(\ref{eq:fNL}) together with the measured amplitude of the scalar perturbations ${\cal P}_{\zeta,\,{\mathrm V}}\simeq 2\times 10^{-9}$ to obtain
\begin{align}\label{eq:C_final21}
\left({\cal C}_{\Omega\zeta}\right)_{\rm S}
\simeq 600\,\frac{H_0^2}{k^2}\,(f_{\rm NL}^{\rm equil})^{1/3}\,e^{-4\pi(\xi_{\rm INT}-\xi_k)}\,\frac{\xi_{\rm INT}^6}{\xi_k^{6}}\,,
\end{align}
which despite the ${\cal O}(10^3)$ coefficient in front, and assuming the factor $\frac{\bar{k}^2}{k^2}\,(f_{\rm NL}^{\rm equil})^{1/3}$ to be of the order of the unity, is exponentially small. For instance, assuming $\xi_k\simeq 2.5$ (which is the largest value of $\xi_k$ allowed by non-observation of nongaussianities in the CMB) and $\xi_{\rm INT}\simeq 5$, which is on the lower end of the values found in numerical studies for $\xi$ in the strong backreaction regime, the factor $e^{-4\pi(\xi_{\rm INT}-\xi_k)}\,\frac{\xi_{\rm INT}^6}{\xi_k^{6}}$ evaluates to approximately $10^{-11}$, making this dimensionless, normalized correlator tiny.

\subsubsection{$\xi_{\rm BR}-\xi_k\gtrsim \frac{N_k-N_{\rm BR}}{2\pi}$}

In this case the integral is dominated by the scales that left the horizon when $\xi$ attained its largest value at the beginning of the strong backreaction regime. Since we are interested in largest value of the momenta, we consider only wavenumbers that reentered the horizon during radiation domination.  The integral   
\begin{align}
\int d\bp\,d\bq	\frac{{\cal A}(\bq,\bk-\bp-\bq,\bp+\bq)\, 	q^{1/2}\,|\bk-\bp-\bq|^{1/2}\,|\bp+\bq|^{1/2}}{(\sqrt{q}+\sqrt{\bk-\bp-\bq})^7\,(\sqrt{q}+\sqrt{\bp+\bq})^7\,(\sqrt{|\bp+\bq|}+\sqrt{|\bk-\bp-\bq})^7}\,,
\end{align}
is estimated in the Appendix, and it evaluates to ${\cal O}(10^{-2})\,e^{6\pi\,\xi_{\rm BR}}/k_{\rm BR}^3$. As a consequence we obtain the result
\begin{align}
&\left({\cal C}_{\Omega\zeta}\right)_{\rm S}(k)\simeq {\cal O}(10^{-2})\frac{k^3}{k_{\rm BR}^3}\frac{H}{f}\frac{e^{2\pi\,\xi_{\rm BR}}}{\xi_{\rm BR}^4}.
\end{align}

In this case the correlator contains an exponentially large factor (for typical values of $\xi_{\rm BR}\approx 5$, one has $e^{2\pi\,\xi_{\rm BR}}={\cal O}(10^{13})$) that is however suppressed by a volume factor ${k^3}/{k_{\rm BR}^3}$ equal to the inverse of the number of patches of size $\sim k^{-1}$. Given that typically strong backreaction kicks in only $\approx 10$ efoldings before the end of inflation (see however~\cite{Garcia-Bellido:2023ser}, where this occurs as early as $\approx 40$ efoldings before the end of inflation), the suppression factor is typically of the order of $e^{-150}\approx 10^{-65}$ (!), making this correlator, also in this regime, tiny.

%%%%%%%%%%%%%%%%%%%%%%%%%%%%%%%%%%%%%%%%%%%%%%%%%%%%%%%%%%%%%%
\section{Discussion and conclusions}
\label{sec:conclusion}%
%%%%%%%%%%%%%%%%%%%%%%%%%%%%%%%%%%%%%%%%%%%%%%%%%%%%%%%%%%%%%%

An important component of current and future gravitational wave research is the detection and characterization of the stochastic gravitational wave background. This background may originate from astrophysical sources or have a cosmological origin. Specifically, identifying a cosmological gravitational wave background will provide important information about the very early universe.

A powerful approach to distinguish between astrophysical and cosmological backgrounds involves studying their anisotropies. Notably, it has been shown that these anisotropies are correlated with the anisotropies in the CMB~\cite{Ricciardone:2021kel,Braglia:2021fxn}. The exploration of such correlations can significantly contribute to the interpretation of the CMB and SGWB measurements.

In the present paper we have investigated the correlator between the curvature perturbation and the energy density of the gravitational waves, computed today, within the axion inflation model. In this model, scalar fluctuations are generated through two distinct mechanisms: first, from the vacuum via the standard amplification process, and second, as a consequence of the production of gauge fields through a process of inverse decay. Consequently, the correlator exhibits two distinct components.

Our analysis shows that the dominant contribution is provided by the correlator with the amplified vacuum fluctuations of the inflaton, that we examined in Section~\ref{subsec:amplified}. Our main result, eq.~(\ref{eq:fin_vacuum}), shows that the normalized correlator between $\Omega_{\rm GW}$ and $\zeta$ could be as large as ${\cal O}(10^{-2})$. The formalism of~\cite{Alba:2015cms,Contaldi:2016koz,Bartolo:2019yeu} can then be applied to derive potentially observable quantities. The actual observability of such correlators, subject to instrumental noise as well as to the intrinsic variance of the isotropic component~\cite{Mentasti:2023icu,Cui:2023dlo}, will depend on the amplitude of the anisotropies in the gravitational wave spectra.  Such an amplitude is encoded in the correlator $\langle\Omega_{\rm GW}(\bx)\,\Omega_{\rm GW}(\by)\rangle$, whose calculation, in the model of axion inflation,  includes the evaluation of the gauge field's eight-point function -- a calculation that we leave to future work (see however~\cite{Bartolo:2019yeu} for work along this direction). Anisotropies might be large. For instance, the lattice study of~\cite{Bethke:2013aba} showed that the spectrum of gravitational waves induced by preheating at the end of inflation display anisotropies with an amplitude of the order of $\sim 10^{-2}$.

\acknowledgments We thank Gabriele Franciolini, Marco Peloso and especially Valerie Domcke for very useful discussions.  This work is partially supported by the US-NSF grant PHY-2112800.

%%%%%%%%%%%%%%%%%%%%%%%%%%%%%%%%%%%%%%%%%%%%%%%%%%%%%%%%%%%%%%%
%%%%%%%%%%%%%%%%%%%%%%%%%%%%%%%%%%%%%%%%%%%%%%%%%%%%%%%%%%%%%%%
\appendix%%%%%%%%%%%%%%%%%%%%%%%%%%%%%%%%%%%%%%%%%%%%%%%%%%%%%%
%%%%%%%%%%%%%%%%%%%%%%%%%%%%%%%%%%%%%%%%%%%%%%%%%%%%%%%%%%%%%%%
%%%%%%%%%%%%%%%%%%%%%%%%%%%%%%%%%%%%%%%%%%%%%%%%%%%%%%%%%%%%%%%

%%%%%%%%%%%%%%%%%%%%%%%%%%%%%%%%%%%%%%%%%%%%%%%%%%%%%%%%%%%%%%%%%%%%%%%%%
\section{Finding the dominant contribution to the integral in eq.~(\ref{eq:int_dpdq})}%%%%%%%%%%%%%%%%%%%%%%%%%%%%%%%%%%%%%%%%%%%%%%%%%
\label{appendix}%%%%%%%%%%%%%%%%%%%%%%%%%%%%%%%%%%%%%%%%%%%%%%%%%%%%%%%%%
%%%%%%%%%%%%%%%%%%%%%%%%%%%%%%%%%%%%%%%%%%%%%%%%%%%%%%%%%%%%%%%%%%%%%%%%%

In this appendix we discuss how to evaluate the integral in eq.~(\ref{eq:int_dpdq})
\begin{align}
\mathcal{I}(\bk,\,\tau)=&\int d\bp\,d\bq \,	e^{2\pi(\xi_1+\xi_2+\xi_3)} \\ \nonumber
&\times\frac{{\cal A}(\bq,\bk-\bp-\bq,\bp+\bq)\, 	q^{1/2}\,|\bk-\bp-\bq|^{1/2}\,|\bp+\bq|^{1/2}}{(\sqrt{q}+\sqrt{\bk-\bp-\bq})^7\,(\sqrt{q}+\sqrt{\bp+\bq})^7\,(\sqrt{|\bp+\bq|}+\sqrt{|\bk-\bp-\bq})^7}\,,
\end{align}
where the quantities $\xi_1$, $\xi_2$ and $\xi_3$ are given in eq.~(\ref{eq:xi123}). 

As discussed in the main body of the paper, the integral $\mathcal{I}(\bk,\,\tau)$ includes a factor (containing inverse powers of $p$ and $q$) that decreases as $p$ and $q$ increase, and a factor $\propto e^{2\pi(\xi_1+\xi_2+\xi_3)}$ that is, on the other hand, an increasing function of those variables. To estimate which contribution dominates the integral we model the function $\xi(\tau)$ as
\begin{align}\label{eq:ansatz_xi}
\xi(\tau)=\left\{
\begin{array}{ll}
\xi_{\rm {BR}}+\delta \log(\tau_{\rm BR}/\tau) &,\,\, \tau<\tau_{\rm BR}\,,\\
\xi_{\rm {BR}}&,\,\, \tau>\tau_{\rm BR}\,,
\end{array}
\right.
\end{align}
where $\tau_{\rm BR}<0$ corresponds to the time when the produced quanta of gauge field start to backreact strongly on the inflating background. This rough modeling of the function $\xi(\tau)$ has the sole purpose of indicating which range of values of $p$ and $q$ dominate the integral in eq.~(\ref{eq:int_dpdq}). Given that in this parameterization  $\xi$ is constant for $\tau>\tau_{\rm BR}$, the integral will receive a subdominant contribution from momenta satisfying $|p\,\tau_{\rm BR}|\gtrsim 1$, $|q\,\tau_{\rm BR}|\gtrsim 1$, so we will limit our integrations to $p,\,q\lesssim 1/|\tau_{\rm BR}|\equiv k_{\rm BR}$. Moreover, since the strong backreaction regime will kick in relatively late during inflation, when the scales that reenter during radiation domination are leaving the horizon, we can set $\hat{\cal T}=$ constant in this regime, and thus ignore the effects of the transfer function in this analysis.

We present here only an analysis of the contribution to $\mathcal{I}(\bk,\,\tau)$ given by the range of momenta where $p\gtrsim k$. We have checked that the contribution from $p\lesssim k$ has no significant effect.	To start with, we estimate the integral in $d\bq$ which is composed by three different relevant momentum intervals 
\begin{equation}
\int d\bq =  \left( \int_{0}^{k} + \int_{k}^{p} + \int_{p}^{k_{\rm BR}}\right) dq \,q^2 \, \int d\Omega_q \,,
\end{equation}
and we subsequently estimate the integrals in $d\bp$, using
\begin{equation}
\int d\bp =   \int_{k}^{k_{\rm BR}}  dp \,p^2 \, \int d\Omega_p \,.
\end{equation}
After performing the integrals in $dq$ and on the solid angles  $d\Omega_p$, $d\Omega_q$, we obtain 
\begin{equation}
\mathcal{I}(\bk,\,\tau)\simeq  \int_{k}^{k_{\rm BR}}  dp \,p^2 (A_1 + A_2 + A_3 )\,,
\end{equation}
with
\begin{align}
&A_1\simeq .9\times\frac{e^{6\pi\xi_{BR}}}{k_{\rm BR}^{6\pi\delta}}\, p^{6\pi\delta-\frac{19}{2}}\, k^{7/2}\,,\nonumber\\
&A_2\simeq .9\times\frac{e^{6\pi\xi_{BR}}}{k_{\rm BR}^{6\pi\delta}}\, p^{6\pi\delta-\frac{19}{2}}\, (p^{7/2}-k^{7/2}) \sim  \,\frac{e^{6\pi\xi_{BR}}}{k_{\rm BR}^{6\pi\delta}}\, p^{6\pi\delta-\frac{19}{2}}\, p^{7/2}\,,\nonumber\\
&A_3\simeq \frac{8\times 10^{-6}}{\delta-1/\pi} \,\frac{e^{6\pi\xi_{BR}}}{k_{\rm BR}^{6\pi\delta}}\,(-p^{6\pi\delta-6}+{k_{\rm BR}}^{6\pi\delta-6}) \sim 
\frac{8\times 10^{-6}}{|\delta-1/\pi|} \,\frac{e^{6\pi\xi_{BR}}}{k_{\rm BR}^{6\pi\delta}}\times
\left\{
\begin{array}{ll}
{k_{\rm BR}}^{6\pi\delta-6}&,\,\, \text{if }\delta> 1/\pi\,,\\
p^{6\pi\delta-6} &,\,\,  \text{if }\delta < 1/\pi\,.\\
\end{array} \right.
\end{align}
Finally, performing the integral on $p$ we have $\mathcal{I}= \mathcal{I}_1 + \mathcal{I}_2 + \mathcal{I}_3$, with
\begin{align}
&\mathcal{I}_1 \simeq\frac{5\times 10^{-2}}{|\delta- 13/(12\pi)|}\,\frac{e^{6\pi\,\xi_{\rm BR}}}{k^3}\times \left\{
\begin{array}{ll}
(k/k_{\rm BR})^{13/2} &,\,\, \text{if } \delta>13/(12\pi)\,,\\
(k/k_{\rm BR})^{6\pi\delta}&,\,\, \text{if } \delta<13/(12\pi) \,,
\end{array} \right.\nonumber\\
&\mathcal{I}_2  \simeq \frac{5\times 10^{-2}}{|\delta- 1/(2\pi) |}\,\frac{e^{6\pi\,\xi_{\rm BR}}}{k^{3}}\times\left\{
\begin{array}{ll}
(k/k_{\rm BR})^3 &,\,\, \text{if } \delta>1/(2\pi) ,\\
(k/k_{\rm BR})^{6\pi\,\delta}&,\,\, \text{if } \delta<1/(2\pi)\,, 
\end{array} \right.\,\nonumber\\
& \mathcal{I}_3  \simeq \frac{3\times 10^{-6}}{|\delta- 1/\pi|}\,\frac{e^{6\pi\,\xi_{\rm BR}}}{k^{3}}\times \left\{
\begin{array}{ll}
(k/k_{\rm BR})^3 &,\,\, \text{if } \delta>1/\pi ,\\
\frac{0.2}{|\delta- 1/(2\pi) |}\,(k/k_{\rm BR})^3&,\,\, \text{if } 1/(2\pi) <\delta<1/\pi, \\
\frac{0.2}{|\delta- 1/(2\pi) |}\,(k/k_{\rm BR})^{6\pi\,\delta}&,\,\, \text{if } \delta<1/(2\pi)\,. \\
\end{array} \right.
\end{align}

In particular, we find that for $\delta<\frac{1}{2\pi}$ the correlator is proportional to the sixth power of the amplitude of the gauge field when the scale $k$  left the horizon, i.e. $e^{\pi(\xi_{\rm BR}-\delta\,\log(k_{\rm BR}/k)}$. On the other hand, for $\delta>\frac{1}{2\pi}$, the result is proportional to the sixth power of the gauge field at the beginning of the strong backreaction regime.
From the definition~(\ref{eq:ansatz_xi}) we deduce that the integral is dominated by the value of $\xi$ when scales $k$ leave the horizon if $\xi_{\rm BR}-\xi_k\lesssim \frac{N_k-N_{\rm BR}}{2\pi}$, it is dominated by the scales that left the horizon at the beginning of the strong backreaction regime. While this result is based on the parameterization~(\ref{eq:ansatz_xi}), we expect it to be generally valid as long as $\xi(\tau)$ monotonically increases during inflation.

\end{document}